\documentstyle[aps,preprint,epsf]{revtex} 
\begin{document}
\preprint{SNUTP/96-052}
\title
{\bf Fermion Ground State of Three Particles 
in a Harmonic Potential Well and Its Anyon Interpolation}

\author{Chaiho Rim\cite{r}}
\address{
Department of Physics, 
Chonbuk National University, 
Chonju,
561-756, Korea.}  
\maketitle

\thispagestyle{empty}    

\medskip


We examine the detail of the  analytic structure of 
an exact analytic solution of three anyons,
which interpolates to the fermion ground state in a harmonic
potential well.
The analysis  is done on the fundamental domain with appropriate 
boundary conditions.
Some remarks on the hard-core conditions
and self-adjointness are made.

\vfil

\newpage

\section{Introduction}

The concept of an anyon \cite{myr1} is 
based on the homotopy group in two space dimensions
and induces the idea of a smooth interpolation  
between bosonic (symmetric) and spinless fermionic (anti-symmetric) 
spectra.
It is believed that the fractional quantum Hall effect  
is an example of the physical phenomenon of anyons \cite{wil1}. 
To realize this theoretically, one usually introduces 
a statistical gauge field of the Aharonov-Bohm type
to the  Schr\"odinger equation for  bosonic (or  fermionic) particles. 
This first quantized scheme may trace back  
\cite{csq} to the equation of motion derived 
from the second quantized Abelian Chern-Simons gauge theory. 

The anyon wavefunctions, however, even for a free system,
have  been hard to get at.
The non-trivial exchange property of particles 
prohibits constructing 
many particle states from one-particle product states.
Due to the discreteness of the energy of the 
system and the various ways to solve the system,
anyonic spectra  in a harmonic potential well are 
widely studied \cite{sol,rim1}. 
However, the solutions obtained up-to-now are reduced to a 
subset of the bosonic or the fermionic full spectra  
when the  statistical parameter, $\alpha$, is put to the  
bosonic ($\alpha=0$) or the fermionic  ($\alpha=1$) limit.
(We restrict $\alpha$ to be
between 0 and 1 without loss of any generality).
Therefore, a missing state problem arises.
Incidentally, the exact solutions obtained analytically so far 
have a linear energy dependence on the statistical parameter.

One might suspect that the missing state phenomenon
is natural since the bosonic limit is supposed to be a singular 
limit in the sense of the Pauli exclusion principle \cite{chorim}.
However, numerical calculations \cite{num} for a few-body system
have demonstrated that 
interpolating states exist between the missing 
solutions and that the energy dependence on the
statistical parameter might be non-linear.
Perturbative investigations \cite{pert} seem to
confirm this non-linear behavior of the spectra
under situations  
lacking exact information about the interpolation.

Recently, a new family of exact analytic solutions
have been  proposed in \cite{rim3}, which supplements the known
exact solutions.
The solutions exhibit a linear energy dependence on the 
statistical parameter and do not satisfy the hard-core condition 
in general. However, the solutions have  not been  fully 
accepted  in the community. 
We therefore, 
present a detailed analysis of the solution, 
which interpolates to the ground state of 
three anyons in  a harmonic potential well
and which corresponds to a typical missing state.

In Section II, we briefly summarize the method to solve the equation.
The explicit form of the solution and its properties are 
given in Section III.
Section IV is the conclusion and discussion.

\section{Radial and harmonic wavefunction}

We summarize how to solve the Schr\"odinger equation 
for three anyons using the  coordinates in 
Refs. 9-11.
Denoting the coordinates of three particles as 
complex numbers, $z_a = x_a + i y_a$
with $a= 1,2,3$,  we have 
the center of mass (CM)  coordinate as
$Z = { (z_1 + z_2 + z_3)\over  \sqrt3}$ and 
the relative motions  (RM) as 
$u={(z_1 + \eta\,  z_2 + \eta^2\, z_3)\over  \sqrt 3}$ and 
$v={(z_1 + \eta^2\, z_2 + \eta\, z_3)\over  \sqrt 3}\,$
where $\eta = e^ {i 2\pi \over 3} $.
The RM can be conveniently parametrized in terms of 
the Euler angles
($\xi$, $\theta$, and $\phi$)
and the scale parameter $r \ge 0$.
That is, 
$u= r w$ and $v = r z$ where
$ w = \sin(\xi/2)\,\, e^{i ( \theta + {\phi  \over 2}) }$ and 
$ z= \cos (\xi/2) \,\, e^{i (\theta -{\phi  \over 2}) }$.

Typically the Euler angles  are defined on $S^3$ or on $SU(2)$, 
and the angles have the ranges $ 0\le  \xi <\pi$, 
$0\le \chi<2\pi $, and  $-2 \pi  \le \psi <2\pi$.
In our case, due to the exchange symmetry, this domain is reduced to 
${S^3 \over  Z_2 \times Z_3} $.
Note that any 
two-particle exchange can be represented in a combination 
of $E$ and $P$, where $E$ is  the second and third 
particle-exchange operation, $(1,2,3) \to (1,3,2)$, 
and $P$ is a  cyclic operation, $(1,2,3) \to (2,3,1)$.  
Referring to the definition of $u$ and $v$, we have 
$P: ( u,v)\to ( \eta^2 u, \eta v) $ and 
$E:  (u, v) \to (v, u)$.
In other words,  we may choose  a fundamental 
domain given as 
$0 \le r$, $ 0\le \xi <{\pi \over 2}$, 
$-{\pi \over 3}\le  \phi<{\pi \over 3}$, and $0 \le \theta <2\pi$
since under $P$ and $E$, we have 
$P :(r, \xi, \phi, \theta) 
\to (r,\xi, \phi+{2\pi\over3}, \theta + \pi)$
and 
$E: (r,\xi, \phi, \theta) 
\to (r, \pi -\xi,  -\phi, \theta)$,
respectively.

The anyon wavefunction is defined to have  a phase $e^{i \alpha \pi}$ 
when any of the two particles are interchanged. 
This requires the wavefunction have the phase under $P$ and $E$
\cite{rim2}
\begin{eqnarray}
E &&:\quad \Psi (r, \pi - \xi,
-\phi, \theta) = e^{i \alpha \pi}
\Psi (r, \xi, \phi, \theta)\,,
\nonumber \\
P &&:\quad \Psi(r, \xi, \phi + {2 \pi \over 3}, \theta + \pi)
= e^{i 2 \alpha \pi}
\Psi(r, \xi, \phi, \theta)\,,
\nonumber \\
EPEP &&: \quad
\Psi(r,\xi, \phi, \theta+ 2\pi) 
= e^{i 6 \alpha \pi} \Psi(r, \xi, \phi, \theta)\,.
\label{EPR}
\end{eqnarray}

The interaction energy  depends only on the 
scale parameter $r$; $V= V(r)= r^2/ 2$.
This system possesses  simultaneous  eigenstates of $H$, $M$, and $L$,
where $H$ is 
the Hamiltonian for the RM:
\begin{equation} 
H = - {1\over {4r^3}} {\partial\over \partial r} 
(r^3 {\partial \over \partial r} )
+ {1\over {4r^2}} M + V(r)\,.
\label{H}
\end{equation}
$M$ is a Laplacian on $S^3$: 
\begin{equation} 
M = - {4\over \sin \xi} 
{\partial \over {\partial \xi}} \sin \xi  
{\partial \over {\partial \xi}} 
+ {1 \over \sin ^2(\xi /2)}({1 \over 2 i} 
{\partial \over {\partial \theta}} + {1 \over i} 
{\partial  \over {\partial \phi}} )^2
+ {1 \over \cos ^2(\xi/2) }({1 \over 2i} 
{\partial \over {\partial \theta}} - {1 \over i} 
{\partial \over {\partial \phi}} )^2 \,.
\label{M}
\end{equation}
Also $L$ is a  relative angular momentum given by
\begin{equation} 
L = u {\partial \over \partial u} + v { \partial \over \partial v}
- \bar u {\partial \over \partial \bar u } - 
\bar v  {\partial \over \partial \bar v }
={1\over i} {\partial \over {\partial \theta}}\,.
\label{L}
\end{equation}

Simultaneous eigenstates of $H$, $M$, and $L$
are given in a factorized form,
$  \Psi_{E, \mu, l} (r, \xi, \phi, \theta)
 = R_{E, \mu}(r) \,\, \Xi_{\mu, l} (\xi, \phi, \theta)
$
where $E$, 
$\mu (\mu + 2)$, and $l$ 
are 
eigenvalues of $H$, $M$, and $L$,
respectively.
$M$ is a positive semi-definite operator \cite{rim3,rim2},
and its eigenvalue is semi-positive definite with 
$\mu$ being  a non-negative number.
For the case of the  symmetric (bosonic) or 
the anti-symmetric (fermionic) representation, 
$\mu$ is restricted to a non-negative
integer.  This is because the angular momentum $L$ takes on 
integer values. On the other hand, 
for the anyon representation,
$\mu$ can be fractional since 
it interpolates 
bosonic to fermionic states.
However, the generic 
behavior of $\mu$ is  not yet known 
in terms of the statistical parameter $\alpha$.
So far, analytically established 
harmonics have the linear dependence
$\mu = \pm 3 \alpha$ mod integer.
The new family of solutions suggest that 
$\mu \!=  \!3\! \pm \! \alpha$.

$\Xi_{\mu, l}$ is a  harmonic on $S^3$, and 
all the statistical information is to be encoded on this harmonic
since  the exchange property in Eq.~(\ref{EPR}) is independent 
of the scale parameter $r$.
Its two-particle analogue is $e^{\pm i\alpha \theta_{12}}$
where $0 \le \theta_{12} \le \pi $ is the angle of the relative 
coordinate. 
The radial part of the solution can be trivially attained
when the harmonics are given. 
It is given in terms of generalized Laguerre function, and 
its energy is given as 
$E= \mu +2$. 
In the following, we will concentrate on the analysis of 
the harmonics only.
In fact, the same harmonic can be used for obtaining 
wavefunctions for other problems 
when the potential energy is scale dependent only.
A trivial example is the free anyon case.

\section{Harmonic with $\mu = 3 -\alpha$}

The state 
with ($\mu\!= \!3\! -\!\alpha,\,\,  l\!=\! -\!3\! +\! 3\alpha)$ 
interpolates between the fermionic
ground state and a bosonic excited state of the harmonic 
oscillator.
The corresponding 
harmonic should contain a homogeneous power of $z$ and $w$'s
since it is an eigenstate of $L$. 
The solutions are multi-valued,  and 
some subtle points can arise. 
To investigate the analytic structure reliably, we 
work on the fundamental domain
as given  in Section II{}.
The exchange property in Eq. (\ref{EPR})
is to be imposed in the form of boundary conditions
\cite{myr2,rim2}:
\begin{eqnarray}
&&\Xi(\xi, \phi, \theta=2\pi) = 
e^{i6\alpha \pi}\,\, \Xi(\xi, \phi, \theta=0)\,,
\nonumber\\
&&\Xi(\xi, \phi={\pi \over 3}, \theta=\pi) = 
e^{i2\alpha \pi}\,\, \Xi(\xi, \phi=-{\pi \over 3}, \theta=0)\,,
\nonumber\\
&&{\partial \over \partial \phi}\,\, \ln \Xi (\xi, \phi={\pi \over 3}, 
\theta=\pi)= {\partial \over \partial \phi} \,\,
\ln \Xi (\xi, \phi=-{\pi \over 3}, \theta=0)\,,
\nonumber\\
&&\Xi(\xi= {\pi \over 2}, \phi, \theta)=
e^{-i\alpha \pi}\,\, \Xi (\xi = {\pi \over 2}, -\phi, \theta)\,,
\nonumber\\
&&{\partial \over \partial \xi}
\,\, \ln\Xi(\xi ={\pi\over 2} ,\phi, \theta) =
-{\partial \over \partial \xi} \,\, 
\ln \Xi (\xi={\pi \over 2}, -\phi, \theta)\,.
\label{B1}
\end{eqnarray}
It should be noted that
the last two identities hold for $0 <\phi \le \pi /3$.
In addition, the harmonic is to be normalized,  and 
any current across the boundary needs to be finite,
which requires 
\begin{equation}
\Xi(\xi=0, \phi, \theta) \,,\quad
\lim_{\phi \to 0} \phi\,
\Xi^*  {\partial \over \partial \xi} \Xi(\xi ={\pi \over 2}, \phi 
 , \theta)\,,\quad
{\partial \over \partial \phi } \Xi(\xi={\pi\over 2},
 \phi =\pm {\pi \over 3}, \theta)
\label{B2}
\end{equation}
be finite.

The harmonic on the fundamental domain is given as \cite{rim3}
\begin{equation}
\Xi_{3-\alpha, -3+3\alpha}(\xi, \phi, \theta)  
= (z^3-w^3)^\alpha\,\,\,
\Phi (z,w)\,.
\label{Xi}
\end{equation}
$\Phi(z,w)$ consists of three terms:
\begin{equation}
\Phi(z,w) = \Phi_0(z,w) + \Phi_1(z,w) + \Phi_2(z,w)
\label{Phi}
\end{equation}
where
\begin{eqnarray}
&&
\Phi_0(z,w) = {\bar z^3 \over (z \bar z)^{2\alpha}}
{(1 + \bar x)^3 
\over 
[(1 + \eta e^{i \pi p} x ) (1 + \eta^2 e^{i\pi p}x)
 (1 + \bar x) (1+ \bar \eta^3 \bar x ) ]^{\alpha}}\,,
\nonumber\\
&&
\Phi_1(z,w) = {\bar z^3 \over (z \bar z)^{2\alpha}}
{(1 + \bar \eta \bar x)^3 
\over 
[(1 + \eta^2  e^{i \pi p} x ) (1 +  e^{i\pi p}x)
 (1 + \bar \eta  \bar x )^2]^{\alpha}}\,,
\nonumber\\
&&
\Phi_2(z,w) = {\bar z^3 \over (z \bar z)^{2\alpha}}
{(1 + \bar \eta^2 \bar x)^3 
\over 
[(1 +  e^{i \pi p} x ) (1 + \eta e^{i\pi p}x)
 (1 + \bar \eta^2  \bar x) (1+ \eta \bar x) ]^{\alpha}}\,,
\label{Phi012}
\end{eqnarray}
and $x = w/z$. 
The factor 
$(z^3-w^3)^\alpha$
satisfies all the anyonic properties given in Eqs.~(\ref{B1})
and (\ref{B2}).
$\Phi(z,w)$, therefore, should satisfy the bosonic boundary condition,
which is given in Eq. (\ref{B1}), 
when $\alpha =0$. 

The first boundary condition in Eq. (\ref{B1}) is trivially satisfied.
To check the second and the third boundary conditions
in Eq. (\ref{B1}), we note that
for $|x| <1$ 
\begin{eqnarray}
&&
\Phi_k( \eta z, \eta^2 w) = \Phi_{k+1} (z,w)\,,
\nonumber\\
&&
\Phi_{3} (z,w) = \Phi_0 (z,w)\,.
\end{eqnarray}
Therefore, one can easily see that at $|\phi| = {\pi \over 3}$
and $0 \le \xi <{\pi \over 2}$
\begin{eqnarray}
&&
\Phi (\xi, \phi={\pi \over 3}, \theta) 
=
\Phi (\xi, \phi= -{\pi \over 3}, \theta)\,, 
\nonumber\\
&&
{\partial \over \partial \phi}
\ln \Phi (\xi, \phi={\pi \over 3}, \theta) 
=
{\partial \over \partial \phi}
\ln \Phi (\xi, \phi= -{\pi \over 3}, \theta)\,. 
\end{eqnarray}

The last two boundary conditions  (when $\xi ={\pi \over 2}$) are very 
delicate to check since there are branch points on this boundary
at $\phi = 0, \pm{\pi \over 3}$. To avoid confusion, we are going to 
work inside the fundamental domain which has no branch points,
and, therefore, we can neglect the multi-valuedness.
The value at the boundary can be reached by taking the appropriate 
limit.
One can prove that for  $ 0 < \tau < {\pi \over 3}$
and $\xi ={\pi \over 2}^-$,
\begin{equation}
\Phi (\xi={\pi \over 2}^-, \phi=-\tau , \theta) 
=
\Phi (\xi={\pi \over 2}^-, \phi= \tau, \theta)\,. 
\end{equation}
More specifically,
\begin{eqnarray}
&&
\Phi_0 (\xi = {\pi \over 2}^-, \phi=-\tau , \theta) 
=
\Phi_0 (\xi={\pi \over 2}^-, \phi= \tau, \theta)\,, 
\nonumber \\
&&
\Phi_1 (\xi = {\pi \over 2}^-, \phi=-\tau , \theta) 
=
\Phi_2 (\xi={\pi \over 2}^-, \phi= \tau, \theta)\,, 
\nonumber \\
&&
\Phi_2 (\xi = {\pi \over 2}^-, \phi=-\tau , \theta) 
=
\Phi_1 (\xi={\pi \over 2}^-, \phi= \tau, \theta)\,. 
\end{eqnarray}
Therefore, all the boundary conditions in Eq. (\ref{B1}) 
are satisfied
except possibly at the branch points, to which we are turning.

At the coincidence limit $x=1$, 
$\Phi_0$ has a smooth limit, 
$\Phi_0 (\xi={\pi \over 2}, \phi=0, \theta) 
= e^{i3 \theta} 2^{3/2} 3^{-1/2}$.
However,  $\Phi_1$ and $\Phi_2$ have a branch cut 
at ($\xi = {\pi \over 2}, \phi= {\pi \over 3}$)
and ($\xi ={\pi \over 2}, \phi= - {\pi \over 3}$), 
respectively, and their 
values are  not determined unambiguously.
If we take the limit from inside the domain, we have 
\begin{equation}
\lim_{\epsilon \to 0}
{\Phi_1(\xi ={\pi \over 2} -\epsilon, \phi = 0, \theta)
\over
\Phi_2(\xi ={\pi \over 2} -\epsilon, \phi = 0, \theta)
}
= 
e^{-i\pi \alpha}
\label{Phiratio1}
\end{equation}
where we neglect the subtleties due to the branch cuts.
If one starts to count multi-valuedness at the branch cuts,
then one can change the ratio in Eq. (\ref{Phiratio1})
to 
\begin{equation}
\lim_{\epsilon \to 0}
{\Phi_1(\xi ={\pi \over 2} -\epsilon, \phi = 0, \theta)
\over
\Phi_2(\xi ={\pi \over 2} -\epsilon, \phi = 0, \theta)
}
= 
e^{-i\pi (1+2n)\alpha}
\label{Phiratio2}
\end{equation}
where n is an integer representing a branch cut.
Therefore, depending on the choice of the branch cut,
the hard-core condition can be met at the coincidence limit for
$\alpha = {q \over p}$ where $q$ and $p$ are coprimes
and odd integers.  This is because 
in this case one can have 
$\Phi_1(\xi= {\pi \over 2}, \phi=0, \theta ) 
+ \Phi_2 (\xi= {\pi \over 2}, \phi=0, \theta) =0$.
Otherwise, $\Phi_1$ and $\Phi_2$ have infinite values at $x=1$,
which is to be canceled by the zero of $(z^3-w^3)^\alpha$ to 
make the harmonic finite.
The choice of the branch cut, however, looks artificial
since the value at the coincidence point is discontinuous
from the ones  at the surrounding points.

Instead, we claim that the hard-core condition 
is not a mandatory one for the anyonic system.
The Hamiltonian can be self-adjoint even if we do not impose  the 
hard-core condition. 
The harmonic given in Eq. (\ref{Xi})
satisfies all the
necessary boundary conditions for the self-adjointness, 
which are given in Eqs.~(\ref{B1}) and (\ref{B2}). 
Similar behavior is observed in a two-particle system \cite{manuel}. 
The two particles are allowed to collide 
when self-adjoint extension is done. 
Here, in the three-particle case, 
two of the three particles are allowed to 
collide whereas  all the particles cannot collide 
simultaneously because of the scale-dependent part.

We may construct the harmonic 
following the series 
expansion in $x$ as described in Ref. 11.
(Note that $\nu$ and $z$ in Ref. 11 correspond to 
$\alpha$ and $x$ in our notation.)
$\mu$ is to be determined by the last two 
boundary conditions in Eq. (\ref{B1}), 
which turns out to be the most difficult part in the
analysis.  We note that this series expansion 
is  equivalent to the approach we use 
in this paper since both approaches allow the small parameter
$x$, neglect the multi-valuedness inside the fundamental domain,
and have the same boundary conditions to be satisfied. 
Since there should be a  unique solution for the system, 
the series expansion will reproduce
the harmonic in Eq. (\ref{Xi}).
In this way, we can conclude that the series expansion gives
$\mu =3-\alpha$.

One might suspect that  the branch cuts
of $\Phi_1$ and $\Phi_2$ at the boundary point 
with $\xi={\pi \over 2}$ and 
$\phi = |{\pi \over 3}|$ may give  an undesirable 
multi-valuedness feature. 
If there is a multi-valuedness contribution,
then this should be interpreted 
as the particle exchange statistical property. 
However, since that the boundary point corresponds to 
the particle configuration where the three particles lie on a 
straight line with particle 2 or particle 3 on the middle of the 
line, a small circular movement of the middle particle 
without enclosing any other particle should 
not give any branch-cut contribution. 

Indeed, one can demonstrate that 
the harmonic has no such branch-cut subtleties.
This is because 
the front factor of the harmonic,
$(z^3 - w^3)^\alpha$ given in Eq. (\ref{Xi}),
takes care of all the particle-exchange properties, 
Eq. (\ref{EPR}). This gives the true branch cuts on the whole domain.
On the other hand,  the rest factor, $\Phi(z,w)$, 
is not going to give any anyonic property. Instead, it gives
a bosonic property such that the whole expression, 
$\Xi$,  has the desired 
exchange properties. 
To obey this, we have only to pass on the {\it value} of 
$\Phi(z,w)$ on the fundamental domain to other domains. 
We recall that $\Phi(z,w)$ is single-valued 
on the fundamental domain.
Therefore,
$\Phi(z,w)$ on the whole domain never has any multi-valuedness.
This is exactly the merit of the harmonic analysis 
on the fundamental domain.
Of course, the analysis is not complete until
one checks the self-adjointness of the system
for the analysis to work without any defect.
The condition for this 
is given in terms of the boundary conditions in 
Eqs.~(\ref{B1}) and (\ref{B2}).
Since the harmonic satisfies all the necessary  boundary conditions,
we can conclude that the harmonic 
has no branch cut at ($\xi={\pi \over 2},
\phi = |{\pi \over 3}|$) nor any of the subtleties thereof.

\section{Conclusion and discussion}

We have presented a detailed analysis for the 
interpolating solution
between the fermion ground state and a bosonic excited state
in a harmonic potential well,
which is known as a missing state. 
This solution has a linear energy  dependence on the statistical 
parameter $\alpha$.

We claim that the hard-core condition is not mandatory
for anyons when two particles coincide with each other.
Instead, one can impose the possibility of two
particles colliding.  Still one can have self-adjointness 
of the Hamiltonian of the system.

The advantage of the analysis on the  fundamental domain is 
clear. It helps us to
avoid unnecessary confusion due to the multi-valuedness of
the harmonics.
In addition, inside the fundamental domain,
one can use a small parameter,  
$w/z$ (the ratio of the two relative coordinates),
to perform a perturbative analysis for a given statistical 
parameter.

The tendency to linear behavior of $\mu$ 
(the Casimir number of the representation) is expected to continue 
not only for a three-anyon system but also for a  many-anyon system.
On the other hand, the issue of smooth interpolation of spectra
needs further investigation. 
This is because the possibility of 
a many-to-one correspondence
of the spectra is not excluded.
We are going to report a detailed analysis 
of this elsewhere.

\center{\bf ACKNOWLEDGEMENT}

This work is supported in part by the Ministry of Education BRSI-2434
and by the  Korea Science and Engineering Foundation
through Science Research Center program of Center 
for Theoretical Physics, Seoul National University 
and project 94-1400-04-01-3.

\end{document}